# Low Resistance Metal Contacts to MoS$_2$ Devices with Nickel-Etched-Graphene Electrodes

Wei Sun Leong,[†] Xin Luo,[‡,¥] Yida Li,[†] Khoong Hong Khoo,[‡,¥] Su Ying Quek,[*,‡] and John T. L. Thong [*,†]

[†]Department of Electrical and Computer Engineering, National University of Singapore, Singapore 117583.

[‡]Department of Physics, Centre for Advanced 2D Materials and Graphene Research Centre, National University of Singapore, Singapore 117546.

[¥]Institute of High Performance Computing, 1 Fusionopolis Way, #16-16 Connexis, Singapore 138632.

*Address correspondence to elettl@nus.edu.sg (experiment); phyqsy@nus.edu.sg (theory)

ABSTRACT  We report an approach to achieve low-resistance contacts to MoS$_2$ transistors with the intrinsic performance of the MoS$_2$ channel preserved. Through a dry transfer technique and a metal-catalyzed graphene treatment process, nickel-etched-graphene electrodes were fabricated on MoS$_2$ that yield contact resistance as low as 200 Ω.μm. The substantial contact enhancement (~2 orders of magnitude) as compared to pure nickel electrodes, is attributed to the much smaller work function of nickel-graphene electrodes, together with the fact that presence of zigzag edges in the treated graphene surface enhances tunneling between nickel and graphene. To this end, the successful fabrication of a clean graphene-MoS$_2$ interface and a low resistance nickel-graphene interface is critical for the experimentally measured low contact resistance. The potential of using graphene as an electrode interlayer demonstrated in this work paves the way towards achieving high performance next-generation transistors.

Two-dimensional transition metal dichalcogenides (TMDs) are materials that present unique electronic and optical properties. Among them, molybdenum disulfide ($MoS_2$) has received the most attention in recent years. Theoretically, $MoS_2$ field-effect transistors (FETs) have been predicted to possess reasonable electron mobility, large transconductance, and superior on/off ratios ($>10^{10}$) due to the large direct bandgap (1.9 eV) for monolayer $MoS_2$.[1-3] More importantly, due to its natural two-dimensional form, $MoS_2$ is expected to be insusceptible to short channel effects endowing it with great promise as an alternate channel material for post-silicon electronics.[1, 3] Despite such potential advantages, in reality, $MoS_2$ devices with such excellent performance have yet to be realized, one of the most serious impediments being the non-trivial contact resistance at the metal-$MoS_2$ interfaces. The large source/drain contact resistance that has been hampering the realization of ideal $MoS_2$ FETs can be attributed to Fermi level pinning in $MoS_2$ close to the conduction band edge as a result of sulfur-vacancy defect level and charge neutral level location.[3, 4] Regardless of this underlying principle, prior attempts to mitigate this by choosing the most appropriate material as contact metallization to $MoS_2$ have met with limited success -- the lowest contact resistance reported to date using scandium contacts is still far from satisfactory.[3] Alternatively, gas[5] and charge transfer based molecular doping[6] on $MoS_2$ flakes have been used to lower the contact resistance of $MoS_2$ devices but accompanying severe degradation in the on/off performance was observed. Recently, chloride molecular doping technique and formation of contacts to metallic-phase $MoS_2$ have been demonstrated to lower the contact resistance of $MoS_2$ transistors to 500 $\Omega.\mu m$[7] and 200-300 $\Omega.\mu m$,[8] respectively. Both techniques involve hours of chemical immersion and how the chemicals affect the $MoS_2$/dielectric interface requires careful investigation for future technology development.

On the other hand, some research groups have attempted the use of graphene as electrical contacts to $MoS_2$ devices since it has the ability to enhance electron injection into the conduction band of $MoS_2$ through lowering of the Schottky barrier height.[9-11] However, graphene by itself as electrode is too thin for electrical characterization (probing) and thus necessitates the use of metal-graphene hybrid as electrical contact for MoS2 devices.[9, 12] The situation at the metal-graphene-$MoS_2$ contact has become more complicated as it combines both metal-graphene and graphene-$MoS_2$ interfaces. Taken individually, the former contact resistance is known to be large (up to hundreds of Ω.mm) without any contact treatment, while cleanliness is a key requirement for the latter interface. Failure to optimize either one of these interfaces could compromise the contact enhancement brought about by the graphene interlayer. For example, the contact resistance for Ti-graphene-contacted MoS$_2$ devices (3.7 ± 0.3 Ω.mm when back gate bias = 30 V)[11] has been reported to be large compared to the lowest contact resistance values reported for MoS$_2$ devices.[8]

Here, we demonstrate that the resistance of metal contacts to semiconducting-phase MoS$_2$ can be as low as that of metallic-phase MoS$_2$ with the use of etched-graphene as buffer layer at the metal-MoS$_2$ interface. To ensure clean graphene-MoS$_2$ interfaces, a polymer-assisted dry transfer technique is used to layer the graphene onto the MoS$_2$. Hence, the bottom side of the exfoliated graphene flake, which has never been exposed to any polymer or solvent, is clean and bonds to the MoS$_2$ crystal lattice through van der Waals force. This is followed by metallization to the graphene using nickel. In order to optimize the metal-graphene interface, we propose to treat the graphene with a nickel-catalyzed etching process prior to stacking on MoS$_2$, which creates in the graphene a vast density of nano-sized pits with zigzag-terminated edges that covalently bond to the deposited nickel (Ni) metallization, thus minimizing the resistance at metal-graphene interface.[13] In brief, the MoS$_2$ devices with Ni-etched-graphene electrodes show not only about 2 orders of magnitude improvement in contact resistance and 3-fold enhancement in effective mobility, but also performance enhancement in terms of on/off current ratio and subthreshold swing.

**Results and Discussion**

The fabrication process of a back-gated MoS$_2$ FET with Ni-etched-graphene sandwiched at metal-MoS$_2$ contacts is illustrated in Figure 1. In order to carry out a fair comparison between MoS$_2$ FETs with and without treated graphene as an interlayer at the source/drain contacts, we fabricated on the same MoS$_2$ strip exfoliated from molybdenite crystal (SPI supplies®) an array of MoS$_2$ back-gated field-effect transistors consisting of both Ni-MoS$_2$ and Ni-treated-graphene-MoS$_2$ contacts (see Methods for details).

Graphene flakes and MoS$_2$ flakes were first exfoliated onto separate oxidized degenerately p-doped silicon substrate with 285 nm thick SiO$_2$ (Figure 1a). Exfoliated bilayer graphene (BLG) was identified and then treated with a Ni-catalyzed etching technique,[13] which involves nickel deposition, annealing in hydrogen, and removal of nickel particles by acid. From the other substrate, a freshly-exfoliated MoS$_2$ strip (16 nm thick) with uniform width (~2 μm) was chosen. Using a polymer-assisted dry transfer technique, the treated graphene was delaminated from its substrate and transferred onto the selected MoS$_2$ strip (Figure 1b). Thereafter, the source/drain contacts were delineated and metallized with 60 nm of Ni (Figure 1c). Finally, the exposed graphene was completely removed by using the Ni electrodes as a self-aligned hard mask in combination with a benign oxygen plasma process (Figure 1d). As can be seen from Figures 2(a) and (b), the exposed graphene was removed by the oxygen plasma process leaving behind only the MoS$_2$ as the effective channel for Ni-treated-graphene contacted MoS$_2$ devices. In addition, the unprotected MoS$_2$ portion shows no visible damage under an optical microscope. Indeed, the output characteristics of Ni-contacted MoS$_2$ devices remain unchanged (Figure 3a) prior to and following the benign oxygen plasma process.

For this study, we extracted the contact resistance using transfer length method (TLM). An array of two-terminal devices with Ni-MoS$_2$ contacts and a corresponding adjacent array of two-terminal devices with Ni-treated-BLG-MoS$_2$ contacts were fabricated on the same MoS$_2$ flake of 2 μm width and channel lengths varying from 0.5 to 3 μm, in steps of 0.5 μm (Figure 2b). The two-terminal resistance of each device was measured as a function of back-gate voltage, $V_G$. Subsequently, the data was fitted using equation (1):

$$R = 2R_C(W) + R_S \frac{L}{W} \qquad (1)$$

where $R$ is the two-terminal resistance of each device being measured, $R_C$ is the contact resistance of each contact as a function of width, $R_S$ is the sheet resistance, and $L$ and $W$ represent channel length and width, respectively. Electrical characterizations on all devices were performed in vacuum at room temperature and no annealing was performed prior to measurement.

The extracted $R_C$ as a function of back-gate voltage is plotted in Figure 3b, with error bars that represent uncertainty in the fitting. The $R_C$ of all devices in this study exhibits clear gate-voltage dependence (Figure 3b). This trend corroborates other findings reported in literature,[6, 14] where the $MoS_2$ is electrically doped under high gate bias, resulting in smaller effective Schottky barrier height, and hence the $R_C$ reduces with back-gate bias. As can be seen in Figure 3b, the fitted $R_C$ for devices with Ni-treated-BLG-$MoS_2$ contacts was found to be 0.26 ± 0.06 Ω.mm and 0.46 ± 0.11 Ω.mm at back-gate biases of 50 V and 0 V, respectively, which is within 15% of the $MoS_2$ FETs' on-resistance and approaches the $R_C$ required for current state-of-the-art silicon MOSFETs.[15] On the other hand, the fitted $R_C$ for Ni-contacted $MoS_2$ devices without any contact interlayer is 9.6 ± 2.1 Ω.mm and 36.6 ± 7.8 Ω.mm at back-gate biases of 50 V and 0 V, respectively, which is about 37 and 80 times higher than the fitted $R_C$ of devices with Ni-treated-BLG-$MoS_2$ contacts, respectively. It should be noted that these 2 sets of devices were fabricated on the same $MoS_2$ strip. For comparison purposes, devices with Ni-single-layer-graphene(SLG)-$MoS_2$ contacts were fabricated on another $MoS_2$ strip of a similar thickness and the fitted $R_C$ is also plotted in Figure 3b. The $R_C$ of devices with Ni-SLG-$MoS_2$ contacts is 0.83 ± 0.18 Ω.mm and 1.80 ± 0.44 Ω.mm at back-gate biases of 50 V and 0 V, respectively, which represents 12 and 20 times improvement, respectively, compared to that of pure Ni contacts to $MoS_2$. On the other hand, with untreated BLG as the contact interlayer, the extracted $R_C$ of 0.81 ± 0.16 Ω.mm and 1.31 ± 0.41 Ω.mm at back-gate biases of 50 V and 0 V, respectively, is comparable to that of Ni-SLG-$MoS_2$ contacts (Figure 3b). Nonetheless, these values are at least 3 times larger than those of Ni-treated-BLG-$MoS_2$ contacts regardless of back-gate biases. This additional contact enhancement with the use of treated BLG as the contact interlayer, compared to both untreated SLG and BLG, can be attributed to the much smaller resistance at the Ni-graphene interface as illustrated by Figure S1. In spite of that, the use of treated SLG as the contact interlayer showed inferior contact performance compared to that of untreated SLG (Table S1). This can be attributed to the perforated treated-SLG contact interlayer that no longer prevents direct interaction between Ni and $MoS_2$ atoms.

It should be noted that the Ni-catalyzed etching treatment performed on graphene (either SLG or BLG) is capable of creating a significant amount of zigzag graphene edges (either triangular or hexagonal nano-sized pits)[13] in the graphene surface (see Supporting Information S3 for details). More importantly, these zigzag graphene edges are expected to directly bond to the subsequent nickel metallization in end-contacted geometry,[13] rather than the surface-contacted geometry, and thus results in metal-graphene interfaces with smaller contact resistance ($R_{edge} \ll R_{surface}$) as has been predicted theoretically[16] and proven experimentally.[17, 18] In addition, due to weak van der Waals bonds, the tunneling resistance between graphene layers in BLG is found to be larger compared to that of end-contacted metal-graphene contacts ($R_{edge} \ll R_{interlayer}$)[19] and hence graphite (>2 layers of graphene) should be avoided to minimize the contribution of tunneling resistance between graphene layers. In brief, the results show that insertion of graphene as an interlayer by itself already enhances the carrier injection at the metal-MoS$_2$ contacts, and the use of the Ni-catalyzed etching technique further boosts the carrier injection at the metal-MoS$_2$ contacts.

In Figure 4, we compare the output and transfer characteristics of two MoS$_2$ FETs with different types of contacts: Ni-MoS$_2$ and Ni-treated-BLG-MoS$_2$ contacts. Both FETs were fabricated with the same device dimensions and on the same MoS$_2$ flake. Electrical characterization were performed in vacuum at room temperature. As can be seen from Figure 4(a), the MoS$_2$ devices show ~10 times improvement in terms of on-current at $V_{DS} = 2$ V and $V_G = 50$ V with the insertion of treated-graphene as a sandwich layer at metal-MoS$_2$ contacts. This observation is consistent across all the fabricated FETs regardless of the channel length. In addition, back-gate measurements were also performed on these pairs of MoS$_2$ FETs and both FETs exhibit n-type behaviour with similar threshold voltage. This implies that the treated-graphene sandwich layer at metal-MoS$_2$ contacts induces no doping. Moreover, the typical MoS$_2$ devices with Ni-treated-BLG-MoS$_2$ contacts exhibit larger on/off current ratio ($10^5$) and better subthreshold-swing (3.7 V/decade at $V_{DS} = 0.2$ V) compared to those of Ni-MoS$_2$ contacts ($10^4$ and 6.7 V/decade at $V_{DS} = 0.2$ V) as can be seen from the inset of Figure 4(b). We then extract the extrinsic mobility of these pairs of MoS$_2$ FETs using equation (2):

$$\mu = \frac{1}{C_{ox} \cdot V_{DS}} \cdot \frac{L_{ch}}{W_{ch}} \cdot \frac{\Delta I_{DS}}{\Delta V_G} \qquad (2)$$

where $C_{ox}$ is the gate capacitance ($1.21 \times 10^{-8}$ F/cm$^2$ for 285 nm thick SiO$_2$), $L_{ch}$ and $W_{ch}$ are channel length and width, respectively, $I_{DS}$ is the drain current, $V_{DS}$ is the drain voltage and $V_G$ is the gate voltage. The electron mobility for the typical device with Ni-treated-BLG-MoS$_2$ contacts is 80 cm$^2$/V-s, while for the device with Ni-MoS$_2$ contacts is 27 cm$^2$/V-s. This 3-fold improvement in extrinsic mobility can be attributed to the reduced contact resistance. It is worth noting that both the contact resistance and mobility values reported here can be better by elimination of residual PMMA left over by the lithography and polymer-assisted dry transfer processes.

We now elucidate, using first principles calculations, the essential physics behind the observed smaller contact resistance in Ni-treated-BLG-MoS$_2$ compared to Ni-MoS$_2$ contacts. Since the largest reduction in contact resistance occurs when Ni-MoS$_2$ contacts are replaced by Ni-SLG-MoS$_2$ and Ni-BLG-MoS$_2$ contacts (Figure 3b), we first focus on the effect of inserting SLG and BLG at the Ni-MoS$_2$ contacts. The Schottky Barrier Height (SBH) is a key factor that determines contact resistances in semiconductor MOSFETs, and in particular in the MoS$_2$ transistors considered here.[20] We therefore perform a comparative SBH analysis on Ni-MoS$_2$, Ni-SLG-MoS$_2$, and Ni-BLG-MoS$_2$ interfaces. Figure 5 shows the calculated minority-spin bands and side views of the corresponding optimized atomic structures, while the majority spin bands and top view of the atomic structures are presented in Figure S4. When MoS$_2$ is in direct contact with Ni, the binding energy between them is large and the equilibrium interface distance is only 1.92 Å, smaller than the value of 2.08 Å for graphene chemisorbed on Ni. The strong binding influences the electronic structure, as can be seen in Figure 5a, where the projected band structure on MoS$_2$ (represented by blue dots with the projection weight represented by dot size) is strongly perturbed and bears little resemblance to that of pristine MoS$_2$ (Figure S4). For the Ni-SLG-MoS$_2$ contact, graphene $p_z$-states (represented by magenta dots in Figure 5b) hybridize strongly with the Ni $d$-states, and serve as a buffer layer separating the MoS$_2$ from Ni. Therefore, the projected band structure of MoS$_2$ is similar to that of an isolated MoS$_2$ layer (Figures 5b and S4). In both cases, the Fermi level is closer to the MoS$_2$ conduction band minimum (CBM) than to the valence band maximum (VBM), so that the SBH is extracted by taking the energy difference between the Fermi level of the electrode and the CBM of $MoS_2$ (Figure 6). Remarkably, the extracted SBH is reduced from 0.8 eV to 0.3 eV with the insertion of a graphene layer at the Ni-$MoS_2$ contacts (Figure 6a). The accuracy of our calculated SBH is verified using higher convergence criteria (Table S2). With the insertion of BLG as a contact interlayer at the Ni-$MoS_2$ contacts, Ni states are sufficiently isolated from $MoS_2$ and its adjoining graphene layer (Figure 5). Thus, we can see an intact $MoS_2$ projected band structure and a graphene projected band structure that exhibits clear graphene characteristics (represented by orange dots in Figure 5c), such as the Dirac conical points which are shifted from the K-point to the Γ-point due to Brillouin zone folding. The extracted SBH for Ni-BLG-$MoS_2$ contacts is very similar to that for Ni-SLG-$MoS_2$ contacts (Figure 6a).

Generally, it has been found that the SBH at metal-$MoS_2$ contacts increases as the work function of the contacting metal electrode increases.[20-22] This trend makes sense considering the fact that the SBH at metal-$MoS_2$ contacts is computed as the difference between the Fermi level of the electrode and the CBM of $MoS_2$; in experiments, the SBH is more accurately given by the difference between the Fermi level of the electrode and the defect levels in $MoS_2$ (which are also close to the CBM).[2, 23] In the simplest consideration, in order for the Fermi level not to fall too far into the band gap of $MoS_2$, the ideal electrode would be one whose work function is as close as possible to (or if not, smaller than) the electron affinity of $MoS_2$. In Figure 6a, the SBH from our calculations are plotted against the work functions of different electrodes (*i.e.* Ni, Ni-SLG and Ni-BLG). These plots strongly suggest that the significant reduction of the SBH is closely related to the reduction of the work function of the Ni-graphene electrodes (both Ni-SLG and Ni-BLG electrodes) compared to that of Ni electrodes. The calculated work functions, $q\phi_M$ of the Ni-graphene electrodes (3.64 eV for Ni-SLG and 4.08 eV for Ni-BLG) are smaller than the calculated electron affinity, $q\chi$ of isolated $MoS_2$ (4.33 eV); in contrast, the work function of Ni is significantly higher (5.52 eV). As the work function of Ni is substantially larger than the electron affinity of $MoS_2$, even with partial Fermi level pinning from metal-induced gap states at the strongly hybridized Ni-$MoS_2$ interface,[21] the resulting SBH is still quite large. On the other hand, due to the weak van der Waals interactions between graphene and MoS$_2$, Ni-graphene electrodes interact with MoS$_2$ mainly by charge transfer, which determines the resulting SBH. Specifically, one expects electrons to be transferred from the electrodes to MoS$_2$, resulting in an interface dipole that raises the relative potential at MoS$_2$, giving rise to a small SBH (Figure 6b). By comparing the magnitude of the dipole at the electrode-MoS$_2$ interface, we found that the dipole moment is stronger in the Ni-SLG-MoS$_2$ interfaces compared to that of Ni-BLG-MoS$_2$, causing a larger upward shift of the MoS$_2$ bands. This is why even though the work function of the Ni-SLG electrode is 0.4 eV smaller than that of the Ni-BLG electrode, the SBH is comparable in both Ni-SLG-MoS$_2$ and Ni-BLG-MoS$_2$ interfaces. Figures 5d-f plot the plane-averaged charge difference $\Delta n$ obtained by subtracting the plane-averaged charge density of the composite slab from that of individual components of the slab using the same supercell, *i.e.* $\Delta n = n_{Ni-graphene-MoS_2}(z) - n_{Ni}(z) - n_{graphene}(z) - n_{MoS_2}(z)$. These plots give further evidence that the interaction between MoS$_2$ and the electrode is larger for the bare Ni electrodes than for the Ni-graphene electrodes. On the other hand, the complicated charge density profile at the Ni-graphene interface shows that there is strong interaction between Ni and graphene, which explains the large reduction in work function of the Ni-graphene electrodes.[24]

The above discussion indicates that the large reduction in contact resistance going from Ni-MoS$_2$ interfaces to Ni-SLG-MoS$_2$ and Ni-BLG-MoS$_2$ interfaces arises from the reduction in SBH due to the significantly smaller work functions of the Ni-graphene electrodes (both Ni-SLG and Ni-BLG electrodes). As we have seen in the literature,[20, 21] the computed SBH is expected to be slightly larger than the experimentally extracted SBH, partly due to the neglect of defect levels and to the fact that monolayer MoS$_2$ is considered in the calculations. However, the trends predicted by theory have been shown to be robust[20, 21] and we expect that the large change in SBH predicted here also exists in experiment.

Moving on to explain the further reduction in contact resistance going from Ni-BLG-MoS$_2$ to Ni-treated-BLG-MoS$_2$ interfaces, we consider the effects of zigzag graphene edges at the Ni-graphene interface (Figure S5a-b). It is interesting that our computed SBH for Ni-treated-BLG-MoS$_2$ interfaces is very similar to that of Ni-BLG-MoS$_2$ interfaces. Furthermore, by performing nonequilibrium Green's function (NEGF) density functional theory calculations,[25] we find that the contact resistance of zigzag-edge-contacted Ni-graphene interface (233 Ω.µm) is significantly smaller than that of surface-contacted Ni-graphene interface (340 Ω.µm; see Methods for details and Figure S6 for geometries). Similar calculations comparing surface and edge-contacted graphene by other metals have also found edge-contacted metal-graphene interfaces to yield lower contact resistances, due to the stronger metal-graphene coupling at these interfaces.[16] Therefore, we can infer that the effect of Ni-mediated etching treatment on reducing the contact resistance arises primarily from the stronger coupling between Ni and zigzag graphene edges. It is important to note that although stronger coupling can reduce the contact resistance as shown here, tunnelling probabilities depend exponentially on the SBH; therefore, the reduction in SBH going from Ni-$MoS_2$ to Ni-graphene-$MoS_2$ interfaces has a larger effect on the contact resistance than the extent of coupling between $MoS_2$ and the electrodes, which we have not considered here. We also find that the SBH of Ni-treated-SLG-$MoS_2$ interfaces is slightly larger than those of Ni-SLG-$MoS_2$ and Ni-BLG-$MoS_2$ interfaces, but smaller than that of Ni-$MoS_2$ interfaces (Table S2, Figure S5c). These results are consistent with the contact resistances measured in experiment (Table S1).

**Conclusions**

We have shown that the insertion of graphene as a contact interlayer into Ni-$MoS_2$ contacts can significantly reduce the contact resistance. A further treatment of the bilayer graphene surface, namely Ni-mediated etching, can further reduce the contact resistance. Sets of TLM structures consisting of multiple FETs each were fabricated with 4 types of contact structures: Ni-$MoS_2$, Ni-SLG-$MoS_2$, Ni-BLG-$MoS_2$, and Ni-treated-BLG-$MoS_2$ contacts on exfoliated $MoS_2$ of similar thickness. Electrical measurements made on all sets of FETs indicate that the contact resistance of the Ni-SLG-$MoS_2$ FETs is significantly reduced by 20-fold as compared to the Ni-$MoS_2$ FETs, and is further reduced by an additional 3-fold for the FETs with Ni-treated-BLG-$MoS_2$ contacts. This significant contact enhancement results in improved field-effect mobility in the FETs with treated-BLG as a contact interlayer, compared to the more traditional FETs with pure Ni as the contact electrodes (80 $cm^2$/V-s *versus* 27 $cm^2$/V-s). From our I-V measurements, it is clear that regardless of the insertion of untreated or treated graphene at the contacts, a much lower effective Schottky barrier height can be observed. First-principles calculations indicate that the large reduction in contact resistance observed experimentally results from the lower work function in the Ni-graphene electrodes, resulting in a smaller SBH in Ni-graphene-MoS$_2$ contacts. The effect of Ni-mediated etching treatment of bilayer graphene further reduces the contact resistance due to stronger coupling between Ni and the zigzag graphene edges. The significant contact resistance reduction, and enhancement of the on-off ratio, subthreshold-swing and mobility with the use of graphene as the contact interlayer is remarkable considering the fact that there is no systematic method reported to date to improve the contact and adhesion of metal to MoS$_2$. Our findings provide an insight into how the contact resistance at metal-MoS$_2$ contacts can be engineered with the use of graphene as an interlayer. Moreover, the proposed Ni-mediated etching treatment for graphene that further lowers the contact resistance is compatible with semiconductor industry manufacturing technology. Hence, with the approach demonstrated in this work, the use of MoS$_2$ as a mainstream electronic material is brought one step closer to the forefront.

**Methods**

**Fabrication of MoS$_2$ field-effect transistor with Ni-etched-graphene sandwiched at metal-MoS$_2$ contacts.** MoS$_2$ and graphene flakes were first exfoliated on separate oxidized degenerately p-doped silicon substrate with 285 nm thick SiO$_2$ (Figure 1a). The exfoliated bilayer graphene was then treated with a Ni-catalyzed etching technique.[13] This started with the deposition of a thin film (2 nm) of Ni on the exfoliated graphene *via* thermal evaporation at a rate of 0.1 nm per second. Subsequently, the graphene sample was annealed at 580 °C for 10 min in a hydrogen ambient and a significant amount of etched pits with well-defined zigzag edges is created on the graphene basal plane based on a Ni-catalyzed gasification process: C (solid) + 2H$_2$ (gas) -> CH$_4$ (gas).[13] Thereafter, all nickel particles were removed using acid. This was followed by spin-coating of a polymethylmethacrylate (PMMA) layer (600 nm thick) and baked at 120°C in an oven for 10 min. The PMMA film with the graphene was then delaminated from the substrate using a Scotch tape and transferred onto the targeted substrate with the freshly-exfoliated MoS$_2$ strip (16 nm thick) of uniform width (~2 µm) (Figure 1b). For this dry transfer technique, a micromanipulator was used to align the treated-graphene to the exfoliated $MoS_2$ (Figure 1c). The sample was then baked at 100 °C to improve adhesion. Subsequently, the source/drain contacts were delineated and metallized with 60 nm of Ni (Figure 1c). Finally, the exposed graphene was completely removed by utilizing the Ni electrodes as a self-aligned hard mask in combination with a benign oxygen plasma process (10 W RF power, 30 V substrate bias, for 30 seconds). Electrical characterization on all devices were performed at room temperature in a high vacuum chamber ($10^{-6}$ mbar) to avoid unnecessary interaction with moisture in ambient.[26]

**First principles calculations.** The first principles calculations are performed using a plane wave basis set at the level of density functional theory (DFT), as implemented in the VASP code.[27] The local spin density approximation (LSDA)[28] is employed for the exchange-correlation functional and projector-augmented wave (PAW)[29] potentials are used throughout the calculation. The (111) surface of Ni has similar lattice constants with graphene. To construct a commensurate interface between Ni-graphene/Ni and $MoS_2$, we use a rotated 30º 6 × 6 supercell of graphene to accommodate the $MoS_2$ 4 × 4 supercell, and the Ni electrode is modeled by a slab consisting of four atomic layers of Ni. The constructed structures have about 1% compressive strain in graphene and such supercells have been observed experimentally.[30] A plane wave kinetic energy cut-off of 450 eV is used for the wave function basis set, and a Gamma centered 3 × 3 × 1 k-point grid is sampled in the Brillouin zone (BZ). Increasing the cut-off energy and k-point grid only leads to minor differences with the present results (Table S2). Dipole corrections are applied in the self-consistent calculations to prevent interactions between moments of different supercells. The slabs are separated by a vacuum of at least 12 Å to prevent interactions between periodic image slabs. The interface structures and internal coordinates are fully relaxed with fixed lattice constants such that the maximum component of the Hellmann−Feynman force acting on each ion is less than 0.02 eV/Å. In the work function calculations, the unit cells of Ni, Ni-SLG and Ni-BLG are used together with a much denser BZ k-point sampling of 44 × 44 × 1. The energy convergence threshold is set at $10^{-6}$ eV.

It is worth noting that the computed work functions in this work are consistent with the values in prior experimental works. The calculated work function of our Ni and Ni-SLG electrodes are respectively 5.52 eV and 3.64 eV, consistent with previous DFT calculations,[31] and also close to the experimental values (5.35 eV for Ni and 3.9 eV for Ni-SLG).[32, 33]

NEGF-DFT calculations were performed using the ATOMISTIX TOOLKIT (ATK) package, that employs a numerical localized basis sets and the Nonequilibrium Green's function (NEGF) formalism. Self-consistent calculations were performed using a double-$\xi$ polarized basis set and a density mesh cutoff of 300 Ry. Exchange correlation was treated using the local spin density approximation[28] and all atoms were relaxed such that the force per atom was less than 0.05 eV/Å. The k-point samplings of 15 x 1 and 15 x 4 were employed for the surface- and edge-contacted interfaces respectively along the in-plane directions. To obtain the transmission spectrum, a denser k-point grid of 500 x 1 and 500 x 50 was used for the side and edge contact interfaces respectively along the in-plane directions. Figure S6 shows the geometries. The contact resistance was obtained from the calculated transmission spectra $T(E)$ using the formula

$$Rc = \left( \int T(E) \frac{e^{(E-E_F)/k_B T}}{\left(e^{(E-E_F)/k_B T}+1\right)^2} \frac{dE}{k_B T} \right)^{-1}$$

with the temperature set to 300K.

*Acknowledgements:* This project is supported by grant R-263-000-A76-750 from the Faculty of Engineering, NUS and grant NRF-NRFF2013-07 from the National Research Foundation, Singapore.

FIGURES

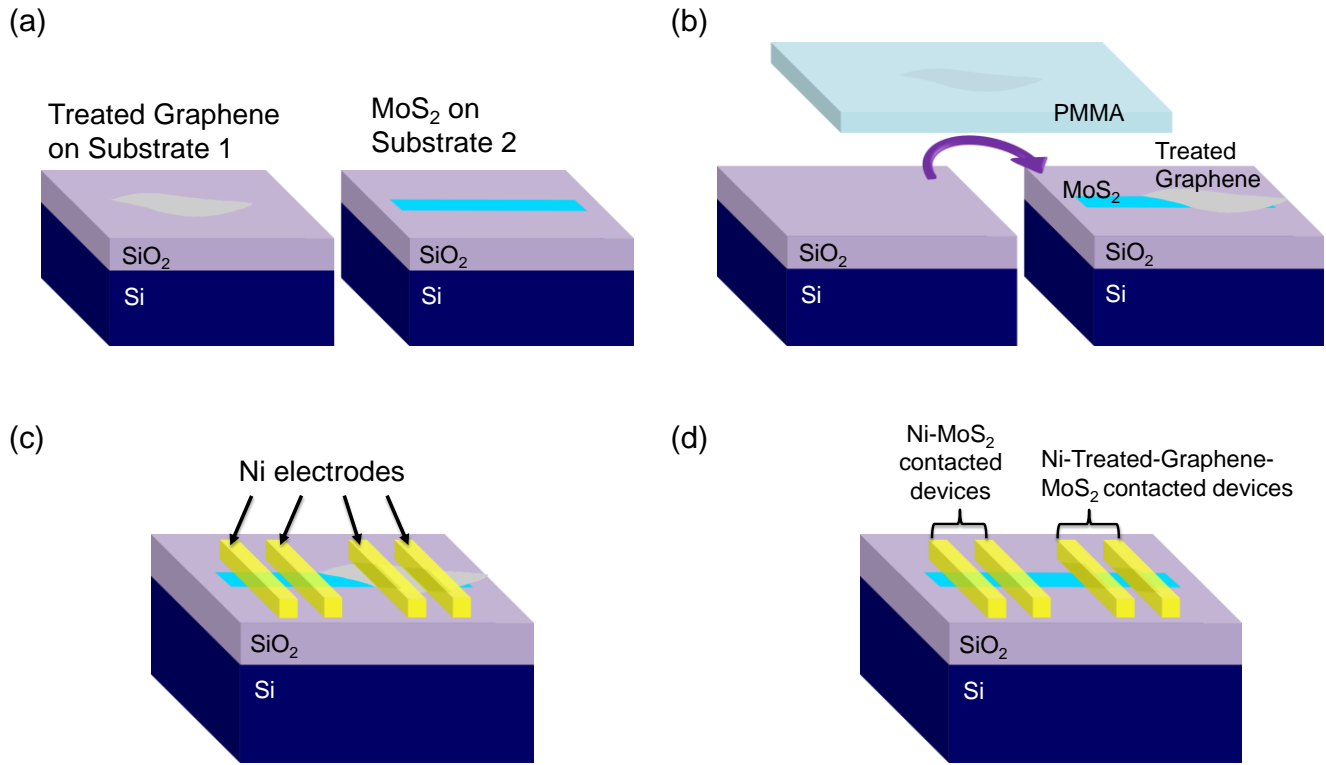

**Figure 1.** Schematics of the process showing the fabrication steps of a back-gated $MoS_2$ field-effect transistor with Ni-etched-graphene sandwiched at metal-$MoS_2$ contacts. (a) Exfoliated graphene on a $p^+$ Si/SiO$_2$ substrate was first treated with Ni-mediated etching process to create large amount of zigzag edges[13] on the graphene surface while pristine $MoS_2$ strip was exfoliated on another $p^+$ Si/SiO$_2$ substrate. (b) The treated graphene was peeled-off from the substrate using "Scotch-tape" technique and transferred onto another substrate with freshly-exfoliated $MoS_2$ strip on it. For comparison purposes, the treated graphene was aligned to cover half of the $MoS_2$ strip. (c) Ni metallization deposited as electrical contacts to the $MoS_2$ device forming both Ni-$MoS_2$ and Ni-treated-graphene-$MoS_2$ contacts. (d) Exposed graphene was removed completely by utilizing the Ni electrodes as self-aligned hard mask in combination with a benign oxygen plasma process.

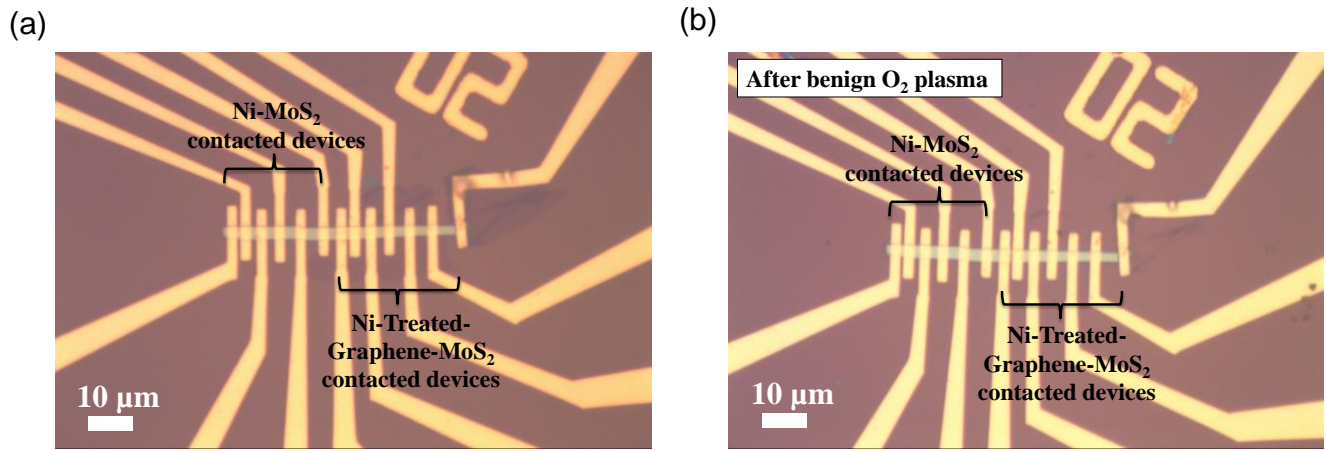

**Figure 2.** Optical images showing a MoS₂ field-effect transistors array consisting of 5 Ni-contacted devices and 6 Ni-treated-graphene-contacted devices with channel lengths varying from 0.5 to 3 µm, in steps of 0.5 µm before (a) and after (b) removal of the exposed graphene by benign oxygen plasma.

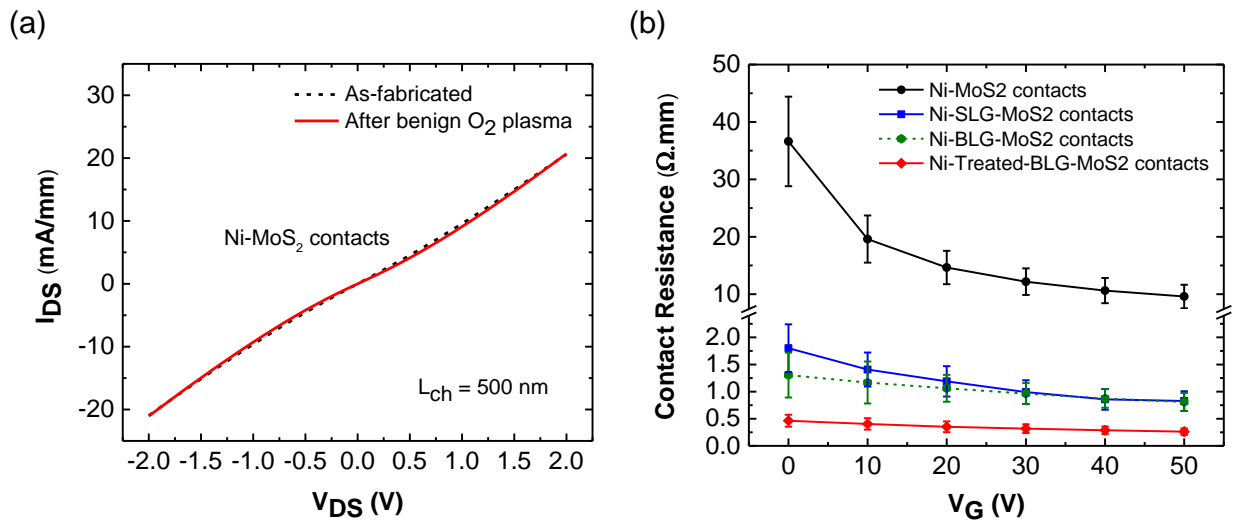

**Figure 3.** (a) $I_D$-$V_D$ characteristics of a typical Ni-contacted MoS$_2$ device shows negligible changes before and after the benign oxygen plasma process. (b) Contact resistance comparison for MoS$_2$ devices fabricated with 4 different types of contacts: Ni-MoS$_2$, Ni-single-layer-graphene-MoS$_2$, Ni-bilayer-graphene-MoS$_2$ and Ni-treated-bilayer-graphene-MoS$_2$ contacts. The lowest $R_C$ of our Ni-treated-bilayer-graphene-contacted MoS$_2$ devices is as low as 200 Ω.μm, which shows ~40 times improvement compared to the Ni-contacted MoS$_2$ devices.

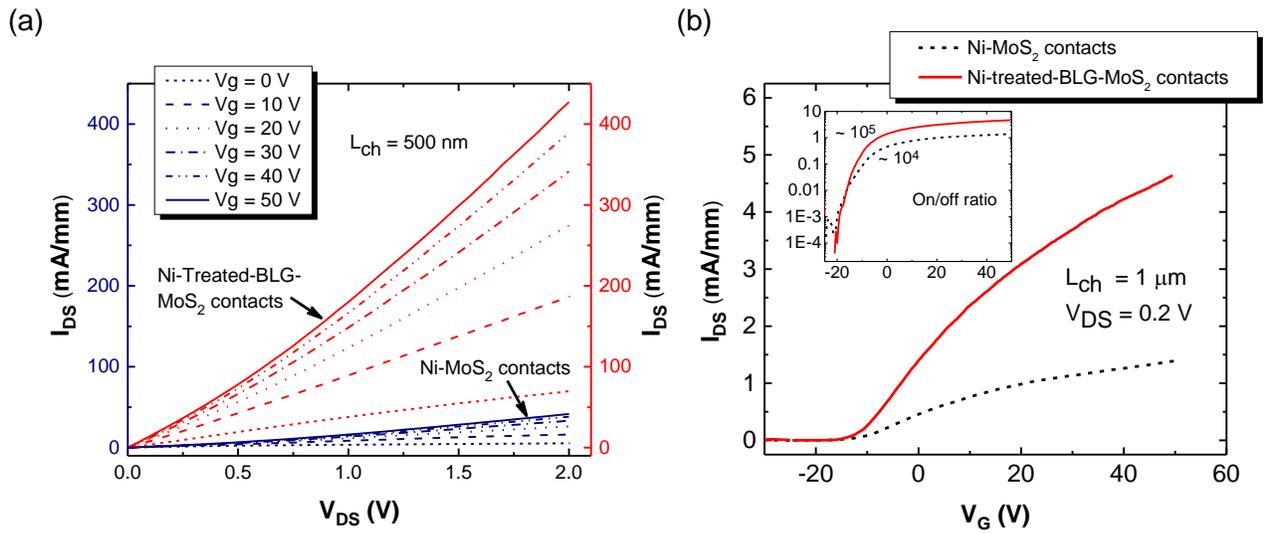

**Figure 4.** (a) $I_D$-$V_D$ characteristics of both $MoS_2$ devices with and without the treated graphene as a sandwich layer. The on-current at $V_{DS}$ = 2 V and $V_G$ = 50 V shows ~10 times improvement, as a result of reduced contact resistance. (b) $I_D$-$V_G$ characteristics of both $MoS_2$ devices with and without the treated graphene as a sandwich layer. Inset: $I_D$-$V_G$ characteristics of the same $MoS_2$ devices in logarithmic scale. Extrinsic mobility of the typical Ni-treated-bi-layer-graphene-contacted $MoS_2$ device is 3 times higher than that of the Ni-contacted $MoS_2$ device.

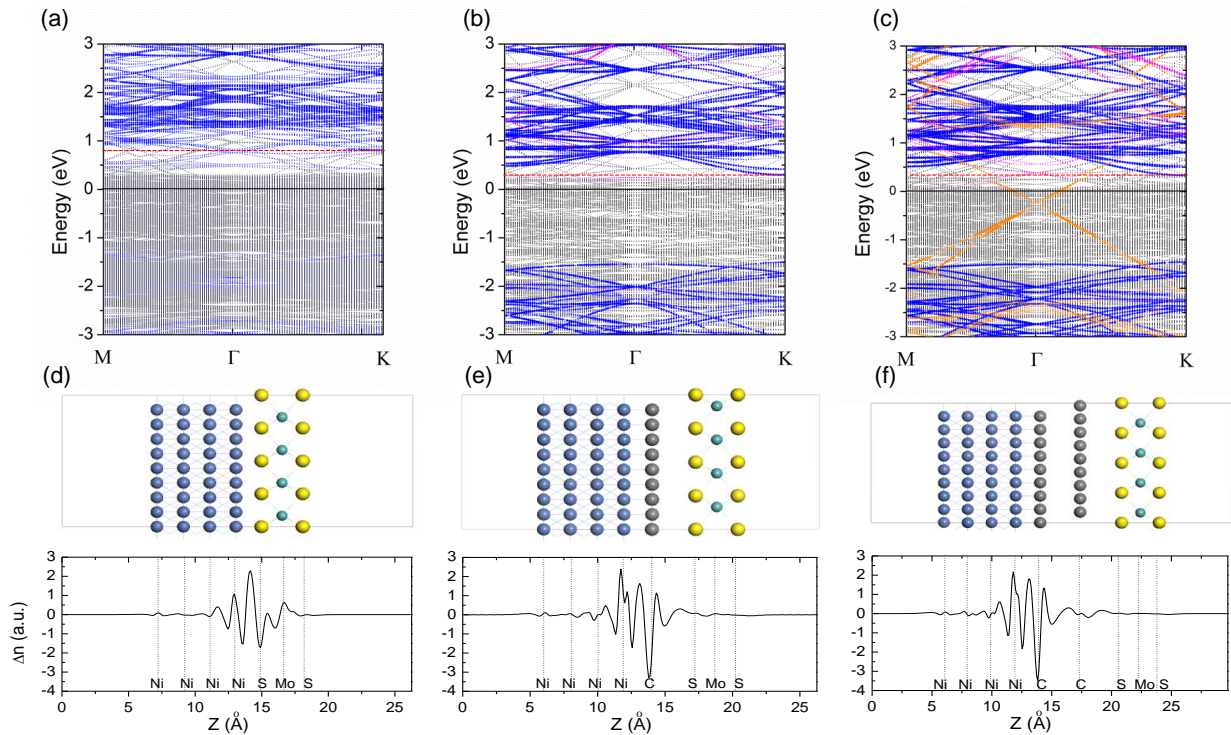

**Figure 5.** Band structures of single layer MoS$_2$ interfaced with (a) Ni, (b) Ni-SLG, and (c) Ni-BLG, only the minority bands are shown here for magnetic systems. The blue, pink and orange dots represent the projected bands of MoS$_2$, graphene layer close to Ni, and graphene layer adjacent to MoS$_2$ respectively, with the projection weight indicated by the dot size. The black solid curves and red dashed curves show the positions of the Fermi level and conduction band minimum of MoS$_2$ respectively. The side views of the atomic structures and plane averaged charge differences Δ$n$ along the z direction are shown below the band structures in panels (d), (e) and (f). The Δ$n$ is obtained by subtracting the plane-averaged charge density of the composite slab from that of individual components of the slab using the same supercell, *i.e.*
$\Delta n = n_{Ni-graphene-MoS_2}(z) - n_{Ni}(z) - n_{graphene}(z) - n_{MoS_2}(z)$.

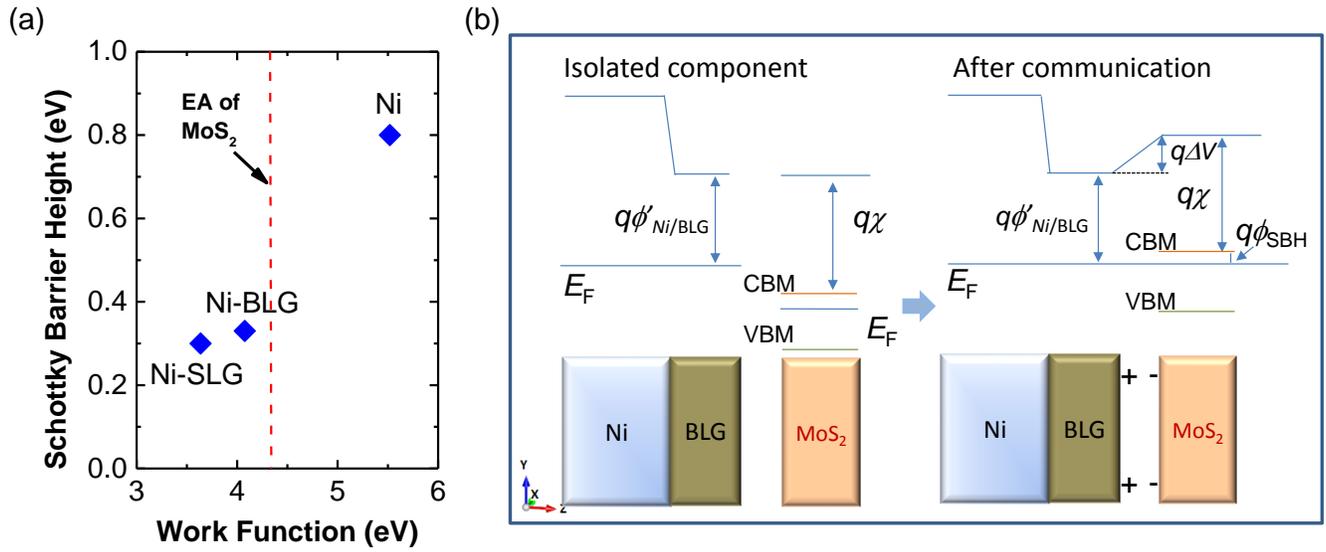

**Figure 6.** (a) Plot of calculated Schottky barrier height as a function of the electrode work function, the electron affinity (EA) of $MoS_2$ is shown in the red dash line. (b) Schematic illustration of the potential shift at the interface of the Ni-BLG-$MoS_2$ system.

**Supporting Information**

# Low Resistance Metal Contacts to MoS$_2$ Devices with Nickel-Etched-Graphene Electrodes

Wei Sun Leong,[†] Xin Luo,[‡,¥] Yida Li,[†] Khoong Hong Khoo,[‡,¥] Su Ying Quek,[*,‡] and

John T. L. Thong [*,†]

[†]Department of Electrical and Computer Engineering, National University of Singapore, Singapore 117583.

[‡]Department of Physics, Centre for Advanced 2D Materials and Graphene Research Centre, National University of Singapore, Singapore 117546.

[¥]Institute of High Performance Computing, 1 Fusionopolis Way, #16-16 Connexis, Singapore 138632.

*Address correspondence to elettl@nus.edu.sg (experiment); phyqsy@nus.edu.sg (theory)

# S1. Average contact resistance on MoS$_2$ devices fabricated using different electrodes

Table S1. The average contact resistance of MoS$_2$ devices fabricated using 5 types of contact structures: Ni-MoS$_2$, Ni-single-layer-graphene(SLG)-MoS$_2$, Ni-bi-layer-graphene(BLG)-MoS$_2$, Ni-treated-SLG-MoS$_2$, and Ni-treated-BLG-MoS$_2$ contacts on exfoliated MoS$_2$ of similar thickness (16 nm).

| Types of electrodes | Number of samples | Contact resistance at back-gate bias = 50 V |
| --- | --- | --- |
| Ni-MoS$_2$ | 5 | 11 ± 3 Ω.mm |
| Ni-SLG-MoS$_2$ | 2 | 0.83 ± 0.18 Ω.mm |
| Ni-BLG-MoS$_2$ | 3 | 0.81 ± 0.16 Ω.mm |
| Ni-treated-SLG-MoS$_2$ | 2 | 1.59 ± 1.08 Ω.mm |
| Ni-treated-BLG-MoS$_2$ | 5 | 0.30 ± 0.10 Ω.mm |

## S2. Comparison of the resistor network models at the metal-MoS$_2$ contacts with bilayer graphene and treated bilayer graphene as a sandwich layer

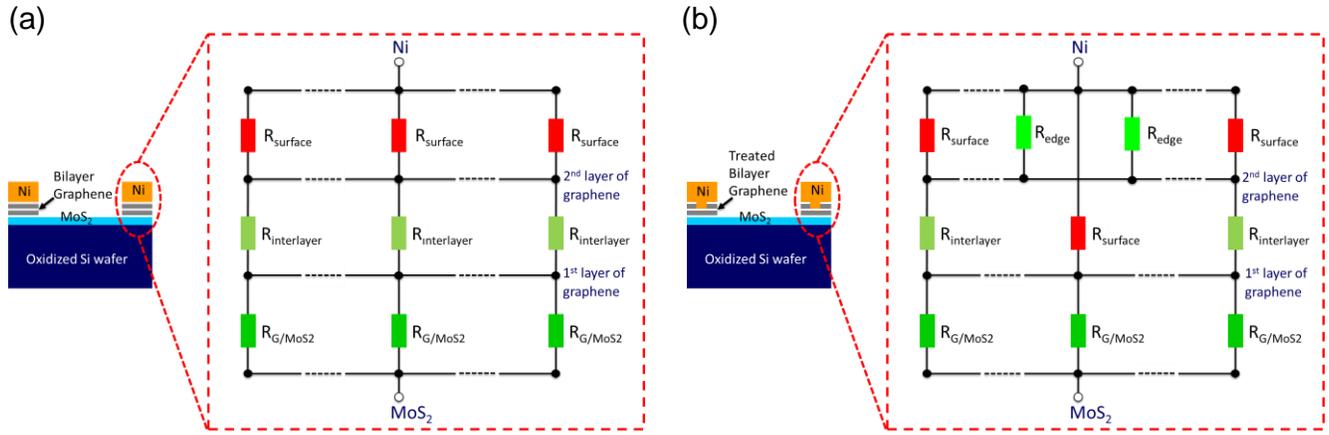

**Figure S1.** Cross-section schematic view of back-gated MoS$_2$ field-effect transistors and resistor network models at the metal-MoS$_2$ contacts with bilayer graphene (a) and treated bilayer graphene (b) as a sandwich layer, where $R_{surface}$ is resistance across the surface-contacted metal-graphene contacts, $R_{interlayer}$ is the tunneling resistance between graphene layers, $R_{G/MoS2}$ is the resistance between graphene and MoS$_2$ interface, $R_{edge}$ is the resistance of the edge-contacted metal-graphene contacts, and particularly $R_{edge} \ll R_{surface}$ and $R_{edge} \ll R_{interlayer}$.

**S3. Characterization of bilayer graphene after the Ni-catalyzed etching treatment**

Here, we characterize a piece of bilayer graphene (BLG) before and after the Ni-catalyzed etching treatment. Using the same recipe as described in main text, we first exfoliate a piece of BLG on a degenerately p-doped silicon substrate with 285 nm thick $SiO_2$ (Figure S2a). The as-exfoliated BLG was then treated with the Ni-catalyzed etching treatment. It was started with the deposition of a thin Ni film (2 nm) on the exfoliated graphene *via* thermal evaporation, followed by annealing at 580 ºC for 10 min in a hydrogen ambient. After that, acid was used to remove the Ni. Figure S2b shows the optical image of the treated BLG. As can be seen, the treated BLG looked similar to when it was freshly-exfoliated although the Ni-catalyzed etching treatment is expected to create a significant amount of etched pits with well-defined zigzag edges on the graphene basal plane as mentioned in the main text.[1] This is because the size of etched pits formed in the graphene surface is in the range of 7 to 27 nm only (Figure S3), which is far beyond the optical resolution limit. In Figure S2c, we compare the Raman spectrum of the BLG when it was freshly-exfoliated and after it undergone the Ni-catalyzed etching treatment. As can be seen, the BLG remains defect-free after the treatment as there is no D-band signal captured at around 1350 $cm^{-1}$, which confirms the etched pits formed in the graphene surface are surrounded by zigzag edges. In addition, the full width at half maximum (FWHM) of the 2D-band signal changes significantly from 50.9 $cm^{-1}$ when it was as-exfoliated to 39.1 $cm^{-1}$ after the Ni-catalyzed etching treatment. The average FWHM of the 2D-band signal of 1600 spectra of the freshly-exfoliated BLG sample is 51.2 ± 1.2 $cm^{-1}$ (Figure S2d), which proves that the graphene is of bilayer.[2] However, the average FWHM of the 2D-band signal of 100 spectra of the treated BLG sample is 37.5 ± 3.4 $cm^{-1}$ (Figure S2d), which is between the range of single-layer (27.5 ± 3.8 $cm^{-1}$) and bilayer graphene (51.7 ± 1.7 $cm^{-1}$) as presented in an earlier Raman studies.[2] Apart from that, the average depth of etched pits is measured to be 0.52 ± 0.13 nm (Figure S3b), while the thickness of the BLG film is about 1.2 nm as measured by tapping mode atomic force microscopy. Overall, the results suggests that the etched pits are most likely to be formed in the uppermost graphene layer only while the bottom layer of the BLG remains as a perfect $sp^2$-hybridized carbon layer.

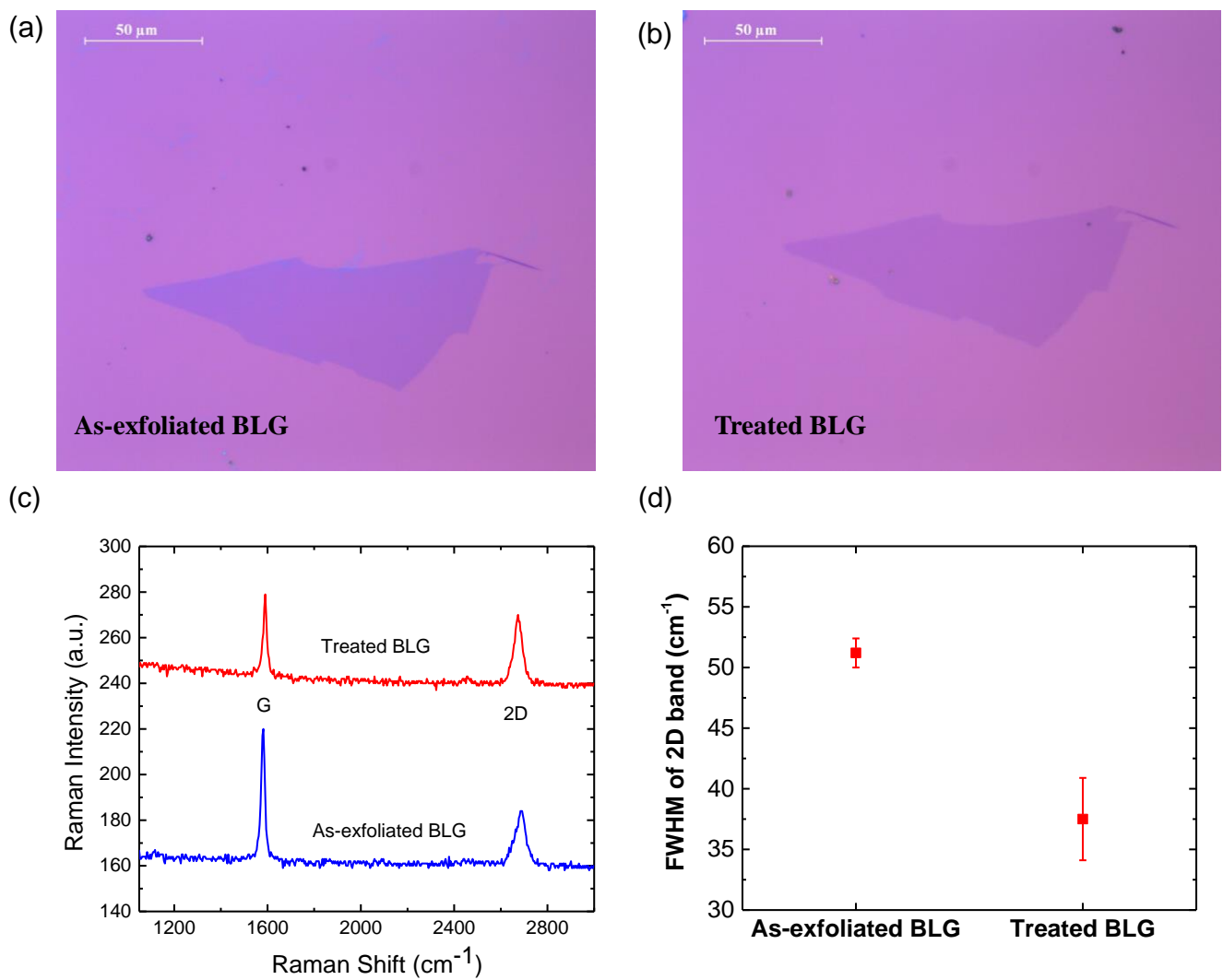

**Figure S2.** Optical image of a bilayer graphene (a) when it was freshly-exfoliated and (b) after the Ni-catalyzed etching treatment. (c) Raman spectrum of the bilayer graphene prior to and following the treatment. (d) The average full width at half maximum (FWHM) of the 2D-band signal of 1600 spectra of the bilayer graphene prior to and following the treatment.

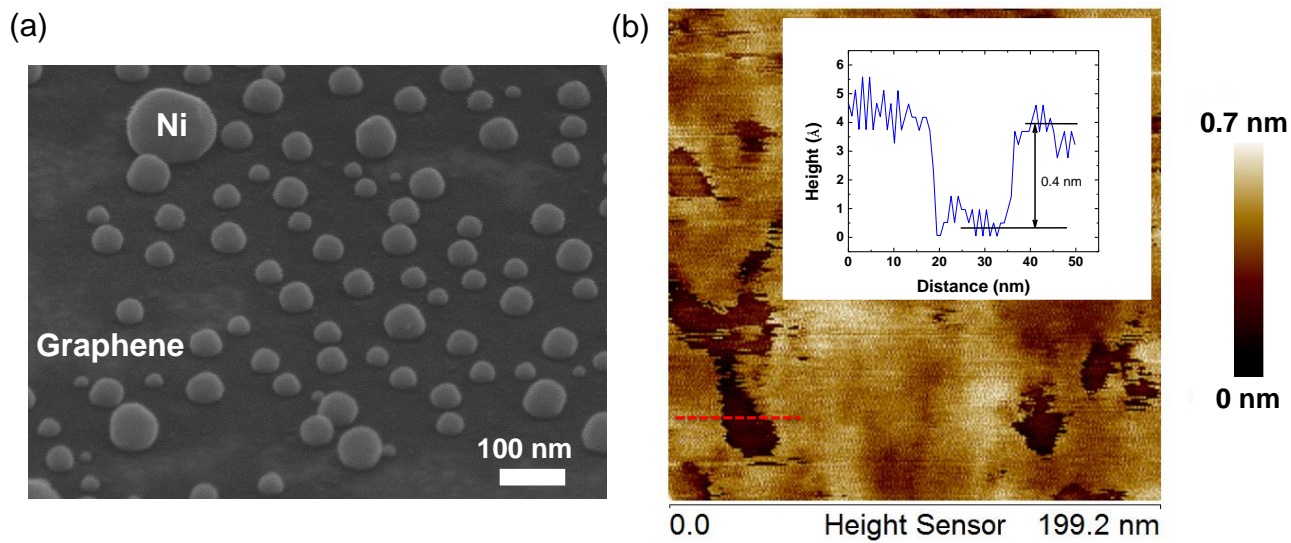

**Figure S3.** (a) Typical 60° tilted SEM image of an etched bilayer graphene showing Ni particles sitting in the middle of each etched pit, which can be partially seen. (b) Typical AFM image of a bilayer graphene after removal of the Ni particles. Inset: Height profile along the dotted line.

## S4. Band structure of MoS₂ interfaced with different electrodes

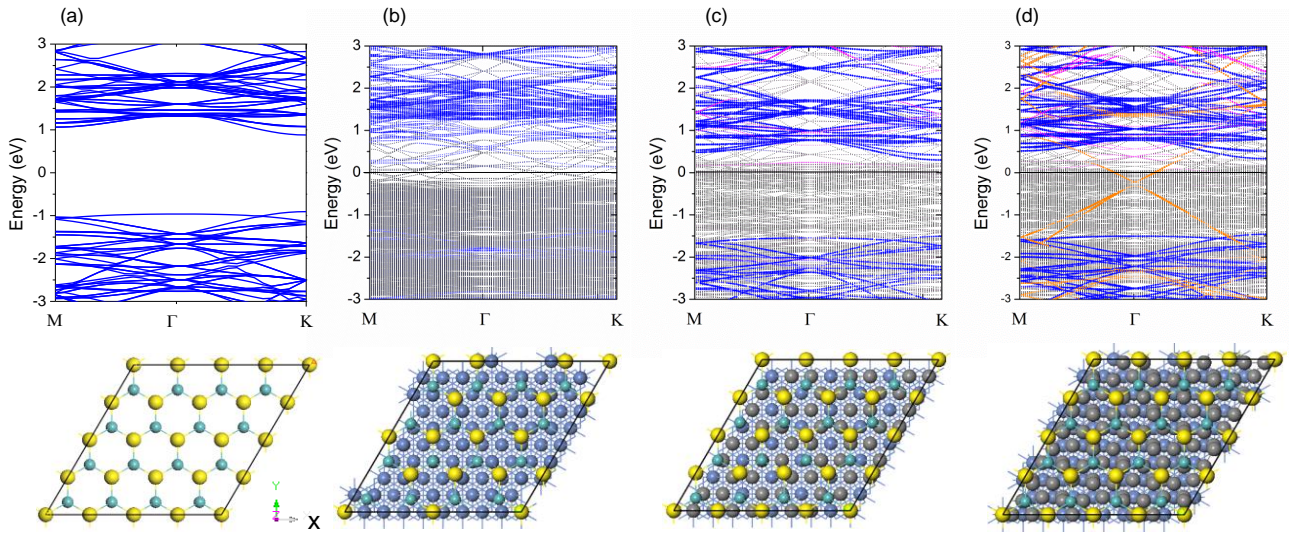

**Figure S4.** (a) Band structure of isolated single layer MoS$_2$ in 4×4×1 supercell. Majority spin band structures for single layer MoS$_2$ interfacing with (b) Ni, (c) Ni-SLG and (d) Ni-BLG, respectively. The top views of their structures are shown at the bottom. The blue, pink and orange dot, respectively, represent the projected band of MoS$_2$, graphene layer close to Ni and MoS$_2$ side, and the weight is indicated by the dot size.

**Table S2**. The calculated Schottky barrier height for different types of interfaces with different k-point sampling of the Brillouin Zone. A smaller supercell is used to obtain the zigzag-edge-contacted Ni-graphene interfaces. Both set of values are very close, indicating the calculation is converged.

| Schottky barrier height / Convergence parameters | k-point: 3×3×1 | k-point: 4×4×1 |
|---|---|---|
| Ni-MoS$_2$ | 0.804 eV | 0.806 eV |
| Ni-SLG-MoS$_2$ | 0.299 eV | 0.290 eV |
| Ni-BLG-MoS$_2$ | 0.343 eV | 0.332 eV |
| Ni-treated-SLG-MoS$_2$ | 0.375 eV | 0.370 eV |
| Ni-treated-BLG-MoS$_2$ | 0.265 eV | 0.265 eV |

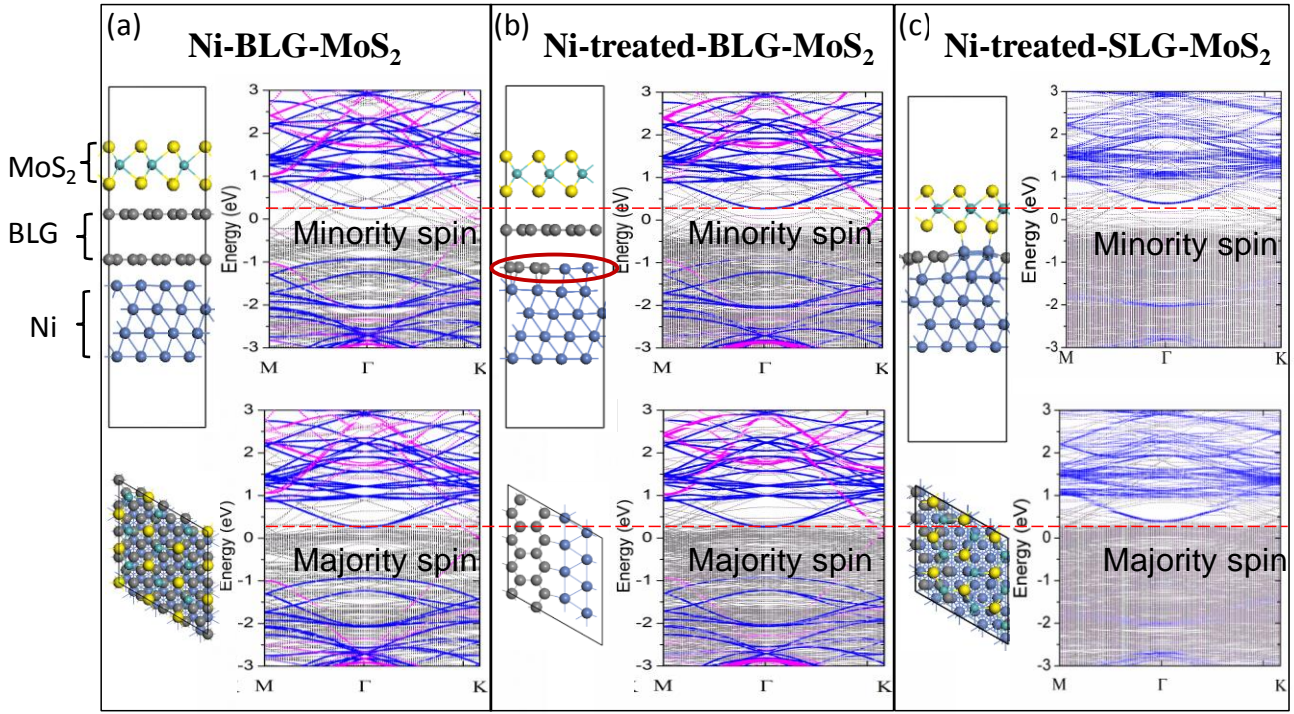

**Figure S5.** Relaxed atomic structures and band structure of (a) Ni-BLG-MoS$_2$ interfaces, (b) Ni-treated-BLG-MoS$_2$ interfaces and (c) Ni-treated-SLG-MoS$_2$ interfaces in 3×3×1 supercell. The blue and pink dots represent the projected band of MoS$_2$ and graphene layer, respectively, and the weight is indicated by the dot size. Both majority and minority spin band structures show a similar Schottky barrier height in Ni-BLG-MoS$_2$ and Ni-treated-BLG-MoS$_2$ interfaces, while the Schottky barrier height of Ni-treated-SLG-MoS$_2$ is 0.11 eV higher than that in the Ni-treated-BLG-MoS$_2$ as shown in Table S2.

## S5. NEGF-DFT calculations of surface-contacted and edge-contacted Ni-graphene interfaces

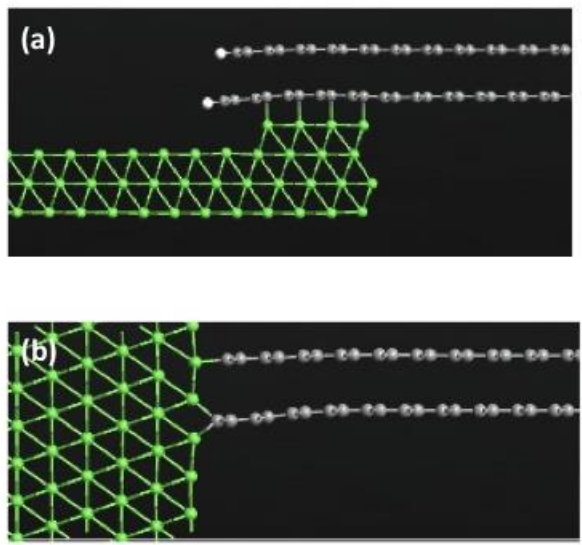

**Figure S6**. Geometries used to model the (a) surface- and (b) edge-contacted Ni-graphene interfaces for NEGF-DFT calculations. Green, grey and white atoms represent Ni, C and H respectively.